\documentstyle[12pt]{article}
\date{}
\newlength{\lll}
\newlength{\lla}
\newcommand{\half}{\mbox{\small$\frac{1}{2}$}}

\newcommand{\calH}{{\cal H}}
\newcommand{\be}{\begin{equation}}
\newcommand{\ee}{\end{equation}}
\newcommand{\bea}{\begin{eqnarray}}
\newcommand{\eea}{\end{eqnarray}}
\newcommand{\mye}{\mbox{e}}
\newcommand{\ph}{\phi}
\newcommand{\bc}{\begin{center}}
\newcommand{\ec}{\end{center}}
\def\rellow#1#2{\mathrel{\mathop{\kern 0pt #1}\limits_{#2}}}
\title{\vbox{\vspace{.1mm}}
Acceptance Rates in Multigrid Monte Carlo}
\author{ \vbox{\vspace{7mm}}
   {\bf Martin Grabenstein$^{1}$ and Klaus Pinn$^{2}$} \\[6mm]
$^1\,$II. Institut f\"ur Theoretische Physik, Universit\"at Hamburg,
      \\[-1mm]
      Luruper Chaussee 149, D-2000 Hamburg 50, Germany\\[4mm]
$^2\,$Institut f\"ur Theoretische Physik I, Universit\"at M\"unster,\\
      [-1mm]
      Wilhelm-Klemm-Str.\ 9, D-4400 M\"unster, Germany\\[4mm]
      }
\begin{document}
\maketitle \vfill
\begin{abstract}
\normalsize
\noindent
An approximation formula is derived
for acceptance rates of nonlocal Metropolis updates
in simulations of lattice field theories.
The predictions of the formula agree quite well
with Monte Carlo simulations of 2-dimensional
Sine Gordon, XY and $\ph^4$ models.
The results are consistent with the following
rule: For a critical model with a fundamental Hamiltonian $\calH(\ph)$
sufficiently high acceptance rates for a complete elimination
of critical slowing down can only be expected if the expansion of
$\langle \calH (\ph + \psi) \rangle$ in terms of the shift $\psi$
contains no relevant term (mass term).
\end{abstract}
\vfill
\thispagestyle{empty}
%%%%%%%%%%%%%%%%%%%%%%%%%%%%%%%%%%%%%%%%%%%%%%%%%%%%%%%%%%%%%%%%%%%%%%%%%
\newpage
\noindent
Monte Carlo simulations of critical or nearly critical
statistical mechanical systems with local algorithms
suffer from critical slowing down.
Roughly speaking, the autocorrelation time in the Markov chain
scales like $\tau \sim \xi^z$, where $\xi$ denotes
the correlation length, and $z$ is the dynamical critical exponent.
For local algorithms, $z \approx 2$.
To overcome this problem, various nonlocal Monte Carlo
algorithms have been developed.

\noindent
Cluster algorithms \cite{cluster} are successful in overcoming
critical slowing down for a large class of models.
The alternative is multigrid Monte Carlo \cite{multigrid,cargmack}.
In this letter, every algorithm that updates stochastic
variables on a hierarchy of length scales is called
multigrid Monte Carlo algorithm.
There are models where no successful cluster algorithms
have been found whereas multigrid Monte Carlo works \cite{sunsun}.
Multigrid Monte Carlo is a candidate for
noncritical simulations of nonabelian lattice gauge theory
\cite{cargmack,thomask}.

\noindent
Presently, the only generally applicable method
to study algorithms for interacting models
is numerical experiment.
It is however important to have some theoretical understanding
that helps to predict which algorithms will have a chance
to overcome critical slowing down
in simulations of a given model.
As a contribution to the research in this direction
we present a study
of the kinematics of multigrid Monte Carlo algorithms.
Sufficiently high
acceptance rates are considered as necessary
for successful multigrid Monte Carlo procedures.
In this letter, we derive and discuss an approximation formula
for the scale (block size) dependence of
acceptance rates for nonlocal Metropolis updates.

\noindent
We consider models with partition functions
\be
Z= \int \prod_{x \in \Lambda_0} d\ph_x \, \exp(-\calH(\ph)) \,
\ee
on cubic $d$-dimensional lattices $\Lambda_0$.
We shall use dimensionless spin variables.
Nonlocal Monte Carlo updates are defined as follows:
Consider the fundamental lattice $\Lambda_0$ as divided in
cubic blocks of size $l^d$. This defines a block lattice
$\Lambda_1$. By iterating this procedure one obtains a whole
hierarchy of block lattices $\Lambda_0, \Lambda_1, \dots, \Lambda_K$.
Let us denote block lattice points in $\Lambda_k$ by $x'$.
Block spins $\Phi_{x'}$ are defined on block lattices
$\Lambda_k$. They are averages of the fundamental field $\ph$
over blocks of side length $L_B=l^k$:
\be
\Phi_{x'} = L^{(d-2)/2}_B \, L^{-d}_B  \sum_{x \in x'} \ph_x \, .
\ee
The $L_B$-dependent factor in front of the average
comes from the fact that the corresponding dimensionful
block spins are measured in units of the
block lattice spacing.
A nonlocal change of the configuration $\ph$ consists
of a shift
\be
\ph_x \rightarrow \ph_x + s \, \psi_x,
\ee
\be\label{normpsi}
L^{-d}_B \sum_{x \in x'} \psi_x = L^{(2-d)/2}_B \delta_{x',x_o'} \, .
\ee
$s$ is a real parameter.
Note that $ \Phi_{x'} \rightarrow \Phi_{x'} + s $
for $x'=x_o'$, and remains unchanged on the other blocks.
The simplest choice of the
``coarse-to-fine interpolation kernel''
$\psi$ that obeys the constraint
(\ref{normpsi}) is a piecewise constant
kernel: $\psi_x = L^{(2-d)/2}_B$, if $x \in x_0'$, and $0$ else.
Other kernels are smooth and thus avoid
large energy costs from the block boundaries. A systematic
study of different kernels will be reported on elsewhere
\cite{tocome}.

\noindent
The $s$-dependent Metropolis acceptance rate for such
proposals is given by
\be\label{omega}
\Omega(s) = \bigl<
\min \lbrack 1 , \exp( - \Delta \calH) \rbrack \bigr> \, ,
\ee
where
\be
\Delta \calH = \calH(\ph + s\psi)-\calH(\ph) \, .
\ee
$\Omega(s)$ has an integral representation
\be\label{intrep}
\Omega(s)= \int du \, \min(1, \mye^{-u}) \int \frac{dp}{2\pi}
\, \mye^{-ipu}
\, \bigl< \mye^{ip \Delta \calH} \bigr> \, .
\ee
Assuming that the probability distribution of $\Delta \calH$
is approximately Gaussian, we are led to
\be
\bigl< \mye^{ip \Delta \calH} \bigr> =
\mye^{ih_1 p - h_2 p^2 } \, \lbrack 1 + o(p^3) \rbrack \, ,
\ee
with
$ h_1 = \bigl< \Delta\calH \bigr> $ and
$h_2 = \half \bigl( \bigl< \Delta\calH^2 \bigr>
     -  \bigl< \Delta\calH \bigr>^2 \bigr) $.
Then the integrations in (\ref{intrep}) can be done exactly:
\be
\Omega(s) \approx
\half \left( \mbox{erfc}(\frac{h_1}{2\sqrt{h_2}})
             + \mye^{h_2-h_1}
              \mbox{erfc}(\frac{2 h_2-h_1}{2\sqrt{h_2}}) \right) \, ,
\ee
\noindent
with
$
{\rm erfc}(x) = \frac{2}{\sqrt{\pi}} \int_x^{\infty}
dt \, \exp(-t^2)
$.
Using invariance properties of the
measure ${\cal D} \ph$ one can show that
the difference of $h_1$ and $h_2$ is
of order $s^4$. We shall see below that the approximation
$h_1 \approx h_2$ is very good.
In this case the acceptance rate prediction simplifies further,
\be\label{formula}
\Omega(s) \approx
\mbox{erfc} ( \half \sqrt{h_1} ) \, .
\ee
(For an analog result in the context of
hybrid Monte Carlo see \cite{hybrid}.)
For free massless field theory with action
$\calH(\ph)= \half(\ph,-\Delta \ph)$,
we obtain the exact result
\be
\Omega(s) = \mbox{erfc}(\sqrt{\frac{\alpha}{8}} \vert s \vert) \, ,
\ee
with $\alpha=(\psi,-\Delta\psi)$. Note that in this case
$h_1=h_2=\half\alpha\, s^2$.
In $d$ dimensions one finds \cite{tocome}
$$
\begin{array}{lcll}
\alpha &=& 2 d L_B \quad &\mbox{for piecewise constant kernels} \, ,
\nonumber \\
\alpha &\rellow{\longrightarrow}{L_B >\!\!> 1}&
\mbox{const} \quad &\mbox{for smooth kernels} \, .
\end{array}
$$
As a consequence, in massless free field theory, to maintain
a constant acceptance rate (of, say, 50 percent) the
changes $s$ have to be scaled down like $L^{-1/2}_B$ for piecewise
constant kernels, whereas for smooth kernels the acceptance
rates do not depend on the block size.
At least for free field
theory, the disadvantage of the piecewise constant kernels can
be compensated for by using a W-cycle instead of a V-cycle.
Smooth kernels can be used only in V-cycle algorithms.

\noindent
In the following we shall apply formula (\ref{formula})
in the discussion of multigrid procedures for
three different spin models in two dimensions:
the Sine Gordon model, the XY model, and the
single-component $\ph^4$ theory.

\noindent
The 2-dimensional Sine Gordon model is defined by the
Hamiltonian
\be
\calH(\ph) = \frac1{2\beta} ( \ph, -\Delta \ph)
- \zeta \sum_x \cos \ph_x  \, .
\ee
The model undergoes a (Kosterlitz-Thouless) phase transition
at $\beta_c$, and $\beta_c \rightarrow 8 \pi$ for $\zeta \rightarrow 0$.
For $\beta > \beta_c$, the flow of the effective Hamiltonian
(in the sense of the block spin renormalization group)
converges to that of a massless free field theory.
The long distance behavior of the theory
is therefore that of a Gaussian model.
One might naively conclude
that multigrid should be the right method to fight
critical slowing down in the simulation of the Sine Gordon model
in the massless phase. But this is not so.
For $h_1$ we find the expression
\be\label{h1h2}
h_1 = \frac{\alpha}{2\beta} s^2
+ \zeta C \sum_x \lbrack  1 - \cos(s\psi_x) \rbrack \, ,
\ee
with $C = \langle \cos \ph_x \rangle$.
The essential point is that the second term in (\ref{h1h2})
is proportional
to the block volume $L^2_B$ for piecewise constant and
for smooth kernels. One therefore has to face a dramatic
decrease of acceptance when the blocks become large, even
for small fugacity $\zeta$. A constant acceptance rate is achieved
only when the proposed steps are scaled down like $L^{-1}_B$.
It is therefore unlikely that any multigrid algorithm
 - based on nonlocal updates of the type discussed in this letter -
will be successful for this model.
We demonstrate the validity of formula (\ref{formula})
(using a Monte Carlo estimate for $C$)
by comparing with
Monte Carlo results at $\beta = 39.478$ and $\zeta=1$.
This point is in the massless phase.
In figure 1  we show both the numerical and analytical results
for $\Omega(s)$ for $L_B=4,8,16,32$ on lattices of size
$16^2,32^2,64^2,128^2$, respectively.

\noindent
We now discuss the
2-dimensional XY model, defined by the partition function
\be
Z = \int {\cal D} \Theta \, \exp \bigl(
\beta \sum_{<x,y>} \cos(\Theta_x-\Theta_y) \bigr) \, .
\ee
As the Sine Gordon model, the XY model has
a massless (spin wave) and a massive phase.
Nonlocal updates are defined by
$ \Theta_x \rightarrow \Theta_x + s \psi_x $, with
$\psi$ obeying again the normalization condition (\ref{normpsi}).
Note that such an update changes
the block spin (defined as the block average of the spins in the
unit vector representation) by an amount $\approx s$ only when
the spins inside the block are sufficiently aligned.
This will be the case in the spin wave phase for
large enough $\beta$.
$h_1$ is given by
\be
h_1 = \beta E \sum_{<x,y>}
\lbrack 1 - \cos( s( \psi_x - \psi_y) )\rbrack \, ,
\ee
with
$E=\langle \cos(\Theta_x - \Theta_y) \rangle $,
$x$ and $y$ nearest neighbors.
For piecewise constant kernels,
$h_1$ is proportional to $L_B$.
For smooth kernels $h_1$ will become independent of $L_B$ for large
enough blocks. For small $s$,
\be
h_1 \approx \half s^2 \beta E \sum_{<x,y>} (\psi_x -\psi_y)^2
= \half s^2 \beta E \alpha \, .
\ee
As above, $\alpha=(\psi,-\Delta \psi)$. This quantity becomes nearly
independent of $L_B$ already for $L_B$ larger than $16$ \cite{tocome}.
{}From the point of view of acceptance rates the XY model therefore
behaves like massless free field theory.
In fact, a dynamical critical exponent $z$ consistent with zero
was observed in the massless phase \cite{xy}.
The failure of multigrid Monte Carlo in the massive
phase ($z \approx 1.4$ for piecewise constant
kernels \cite{xy})
is an example for the fact that good acceptance
rates are not sufficient to overcome critical slowing down.

\noindent
We again checked the accuracy of formula (\ref{formula}) by comparing
with Monte Carlo results at $\beta=1.2$ (which is in the spin
wave phase).
The only numerical input for the analytical formula
was the link expectation value $E$.
The results are displayed in figure 2.
One can do a similar discussion for the $O(N)$ nonlinear
$\sigma$-model with $N>2$, leading to the same prediction
for the scale dependence of the acceptance rates.
This behavior was already observed
in multigrid Monte Carlo simulations of the 2-dimensional
$O(3)$ nonlinear $\sigma$-model with smooth and piecewise constant
kernels \cite{hmm}.

\noindent
Let us finally turn to single-component
$d$-dimensional $\ph^4$ theory, defined
by the action
\be
\calH(\ph) = \half(\ph, -\Delta \ph)
           + \frac{m_o^2}2 \sum_x \ph_x^2
           + \frac{\lambda_o}{4!} \sum_x \ph_x^4 \, .
\ee
For $h_1$ one finds
\be
h_1 = s^2 \, \left\{
\half \alpha + \lbrack \frac{m_o^2}{2} + \frac{\lambda_o}{4}
P \rbrack \sum_x  \psi_x^2 \right\}
+ s^4 \, \frac{\lambda_o}{4!} \sum_x \psi_x^4 \, ,
\ee
where
$P=\langle \ph_x^2 \rangle$.
We have used that expectation values of operators
which are odd in $\ph$ vanish on finite lattices.
Because of the normalization condition (\ref{normpsi}),
$\sum_x \psi_x^2$ increases with $L^2_B$, independent
of $d$, whereas
$\sum_x \psi_x^4$ scales like $L^{4-d}_B$,
for smooth and for piecewise constant kernels.
Therefore also in this model we have to face rapidly
decreasing acceptance rates when the blocks become large.
As in the case of the Sine Gordon model, $s$ is to be
rescaled like $L^{-1}_B$ in order to maintain constant
acceptance rates.
(Note that this discussion is only of relevance when
the correlation length $\xi$ is larger than the block size
$L_B$. The essential point is that the acceptance rates
should not become too small before $L_B$ is of the order of $\xi$.)
{}From our point of view therefore there is little
hope that multigrid algorithms of the type considered
in this letter can overcome critical slowing down
in the 1-component $\ph^4$ model.
In numerical simulations of 2-dimensional $\ph^4$ theory,
a dynamical critical behavior is found
that is consistent with $z \approx 2$
for piecewise constant and for
smooth kernels \cite{phisokal,linn}.
In four dimensions, there is no definite estimate
for $z$ \cite{phifour}.

\noindent
Figure 3 shows a comparison of Monte Carlo results
for 2-dimensional $\ph^4$ theory with the theoretical
prediction based on the numerical evaluation of $P$.
The simulations were done in the symmetric phase
at $m_o^2 = - 0.56$ and
$\lambda_o = 2.4$. The correlation length at this point
is $\xi \approx 15$ \cite{linn}.

\noindent
In this letter, we have presented a simple yet accurate formula
that expresses acceptance rates for nonlocal update algorithms
in terms of one single parameter (entering $h_1$), which is easy
to compute, e.g.\ by Monte Carlo simulations
on a small lattice. We encountered two classes of models.
For Sine Gordon and $\phi^4$ theory $s$ had to be rescaled
like $L^{-1}_B$ for piecewise constant and for smooth kernels, whereas
for the XY model and the $O(N)$ nonlinear $\sigma$-model
one can achieve $L_B$-independent
acceptance rates (choosing smooth kernels).

\noindent
The results are consistent with the following
rule: Sufficiently high acceptance rates for a complete elimination
of critical slowing down can only be expected if
$
h_1 = \langle \calH(\ph+s \psi) - \calH(\ph) \rangle
$
contains no relevant operator in $\psi$. The (superficial)
degree $r$ of
relevance of a local operator is defined by $r=d+m(2-d)/2-n$.
$m$ is the number of fields, and $n$ counts the
derivatives. An operator is called relevant if $r > 0$.
A mass term always has $r=2$, and a $\ph^4$-term has $r=4-d$.
A kinetic term $\alpha=(\psi,-\Delta \psi)$ has $r=0$.
The above statement is formulated for smooth kernels. For
piecewise constant some modifications are necessary. The main
difference is that
$\alpha \propto L$ for piecewise constant kernels.

\noindent
With the help of the rule stated above
it is possible to decide whether a given
statistical model is a natural candidate for multigrid Monte
Carlo or not.
We believe that our formula can be useful
to study different choices of
interpolation kernels $\psi$ in multigrid Monte Carlo for
nonabelian gauge theories \cite{cargmack,thomask}.

\vspace{3mm}
\noindent
{\bf Acknowledgment}

\noindent
We would like to thank G.\ Mack and S.\ Meyer for useful
discussions.
All numerical computations reported on in this paper were
performed on the NEC SX-3 of the University of Cologne
and the CRAY Y-MP of the HLRZ in J\"ulich.
%%%%%%%%%%%%%%%%%%%%%%%%%%%%%%%%%%%%%%%%%%%%%%%%%%%%%%%%%%%%%%%%%%%%%%%%%
\newpage
\mbox{}

\newpage
\bc
{\bf Figure Captions}
\ec
\vskip5mm \mbox{}
\\
FIG.1. $\Omega(s)$ in the 2-dimensional Sine Gordon model,
       $\beta=39.478$, $\zeta=1$.
       From top to bottom: $L_B=4,8,16,32$ on a
       $16^2,32^2,64^2,128^2$ lattice, respectively.
       Points with error bars: Monte Carlo results,
       lines: analytical results.
\vskip1cm
\noindent
FIG.2. $\Omega(s)$ in the 2-dimensional XY model,
       $\beta=1.2$.
       From top to bottom: $L_B=4,8,16$ on a
       $16^2,32^2,64^2$ lattice, respectively.
       Points with error bars: Monte Carlo results,
       lines: analytical results.
\vskip1cm
\noindent
FIG.3. $\Omega(s)$ in the 2-dimensional $\phi^4$ theory,
       $m_o^2=-0.56$, $\lambda_o=2.4$.
       From top to bottom: $L_B=4,8,16$ on a
       $16^2,32^2,64^2$ lattice, respectively.
       Points with error bars: Monte Carlo results,
       lines: analytical results.
\end{document}